\documentclass[conference]{IEEEtran}
\IEEEoverridecommandlockouts
\usepackage{cite}
\usepackage{amsmath,amssymb,amsfonts}
\usepackage{algorithm}
\usepackage{graphicx}
\usepackage{textcomp}
\usepackage[noend]{algpseudocode}
\usepackage{xcolor}
\usepackage{hyperref}
\hypersetup{
  colorlinks   = true, 
  urlcolor     = blue, 
  linkcolor    = black, 
  citecolor   = black 
}
\begin{document}

\pagestyle{plain}

\title{
Literature Survey on Finding Influential Communities in Large Scale Networks}

\author{\IEEEauthorblockN{Prakhar Ganesh}
\IEEEauthorblockA{2015CS10245}
\and
\IEEEauthorblockN{Rahul Agarwal}
\IEEEauthorblockA{2015CS10247}
\and
\IEEEauthorblockN{Saket Dingliwal}
\IEEEauthorblockA{2015CS10254}
}

\maketitle

\begin{abstract}

Community or modular structure is considered to be a significant property of large scale real-world graphs such as social or information networks. Detecting influential clusters or communities in these graphs is a problem of considerable interest as it often accounts for the functionality of the system. We aim to provide a thorough exposition of the topic, including the main elements of the problem, a brief introduction of the existing research for both \textit{disjoint} and \textit{overlapping} community search, the idea of influential communities, its implications and current state of the art and finally provide some insight on possible directions for future research.

\end{abstract}

\section{Introduction}

Many large scale real-world networks like social networks consist of community structures. Disciplines where systems are often represented as graphs, such as sociology, biology and computer science contain examples of such large scale networks and community structures present in them \cite{Li:2015:ICS:2735479.2735484}. The problem of discovering communities in a large scale network is a problem which has attracted much attention in recent years \cite{FORTUNATO201075, Xie:2013:OCD:2501654.2501657}. A similar problem is community search where the goal is to find the most likely community that contains an input query node \cite{Sozio:2010:CPP:1835804.1835923, Cui:2013:OSO:2463676.2463722}. The community discovery problem is focused on identifying all communities present in a network \cite{FORTUNATO201075}. On the other hand, the community search problem is a query-dependent variant of the community discovery problem, which aims to find the community that contains the query node \cite{Sozio:2010:CPP:1835804.1835923}.

An important aspect of communities, which is ignored by the older community detection algorithms, and have been very recently introduced in these field is the influence of a community \cite{Li:2015:ICS:2735479.2735484}. Finding the 'r' most influential communities in a network has widespread applications and has fueled a lot of recent research in this area. The common approaches used in this field includes index based search \cite{Li:2015:ICS:2735479.2735484}, heuristics based on common graph structures like cliques \cite{unknown}, k-cores \cite{Li:2017:FIC:3159176.3159197}, multi-values graphs \cite{Li:2018:SCS:3183713.3183736}, online searches, both global \cite{Li:2015:ICS:2735479.2735484} and local \cite{Bi:2018:OPA:3213880.3232249} and introduction of novel data structures \cite{Li:2017:FIC:3159176.3159197} among many others. We try to cover them all and also provide a quantitative, as well as a qualitative, comparison between the two, along with directions for future research.

\section{Motivation}

The problem of community search over very large graphs is a fundamental problem is graph analysis. However certain applications require us to find the 'r' most influential communities in the graph. Consider the following examples as described in Bi et al. \cite{Bi:2018:OPA:3213880.3232249}. Assume that Alice is a database researcher. She may want to identify the most influential research groups from the co-authorship network of the database community, so as to be aware of the recent trends of database research from those influential groups. Another frequently encountered example is in online social network domain. Suppose that Bob is an online social network user. He may want to follow the most influential groups in the social network, so as to track the recent activities from those influential groups. 

Both of these issues are common in the sense that they need to find communities in a graph with an additional property of having high influence. A possible definition of influence value can be defining it as the minimum weight of the nodes in that community \cite{Li:2015:ICS:2735479.2735484}. An influential community is the one which has a large influence value, which in this case would mean every member of the community is highly influential (since we are taking the minimum).

There has been a sudden outburst of different techniques in this field in the last few years, both in terms of new heuristics in the detection of most influential communities and in terms of speeding up the search algorithms. We aim to provide a thorough survey of the current progress as an entry point for future researchers and insights on possible future research directions.

\section{Problem Statement}

Consider a Graph \begin{math}G(V,E)\end{math} where \textit{V} is the set of vertices and \textit{E} is the set of edges. Let \begin{math}d(v,G)\end{math} be defined as the degree of vertex \textit{v} in graph \textit{G} i.e number of edges coming out of \textit{v}. A graph \begin{math}H(V_h,E_h)\end{math} is defined as an induced subgraph of \textit{G} if \begin{math}V_h \subseteq V \end{math} and \begin{math}E_h = \{ (u,v) : u,v \in V_h, (u,v) \in E \}.\end{math} A k-core is an induced subgraph where the degree of each node is atleast k ( \begin{math}d(v,H) \geq k\end{math}). A maximal k-core is a k-core which is not a subset of any other k-core. Clearly, every graph will have a unique maximal k-core for every values of 'k'.

To introduce the concept of influence we assign a weight to each node which can be interpreted as its influence (such as PageRank, h-index etc). A weight vector of a Graph is the set of all weights assigned to the nodes.

Given a graph \begin{math}G(V,E)\end{math} and an induced graph \begin{math}H(V_h,E_h)\end{math}, we define a function to compute the influence of this induced subgraph. One possible definition, as discussed earlier, could be taking the minimum weight present in the induced subgraph. The intuition behind it is that an induced subgraph is as influential as its least influential member \cite{Li:2015:ICS:2735479.2735484}. Another way could be taking the average of all the weights present in the subgraph or even a more complex function can be defined.

A k-influential community \cite{Li:2015:ICS:2735479.2735484} is an induced subgraph \begin{math}H_k = (V_k,E_k)\end{math} of \textit{G} that :-
\begin{enumerate}
\item Is connected
\item Each node has degree atleast k
\item Is maximal
\end{enumerate}

\textbf{Problem :} Given a graph \begin{math}G = (V, E)\end{math}, a weight vector \textit{W}, and two parameters \textit{k} and \textit{r}, the problem is to find the top-r k-influential communities with the highest influence value. \cite{Li:2015:ICS:2735479.2735484}

\section{Algorithms Used}
The problem statement for finding the most influential communities was first  formulated by Li et al \cite{Li:2015:ICS:2735479.2735484}. They developed a formal definition of influence of an individual and a community, and suggested various methods and optimization to solve the problem. The idea behind their algorithm is to start with the maximal k-core of the graph and then remove nodes from it one at a time, following a certain rule, which will give us the communities present in the graph in an increasing order of influence. The details of the algorithm are discussed below.

\textbf{k-core} : The first step is to calculate the maximal k-core of the graph. This will contain disjoint sets of connected components, each of them satisfying the definition of a community. We then remove a node, present in the k-core, which has the least influence and store its parent community. The maximal k-core of this new graph (formed after removing the node as mentioned above) is calculated and the whole process is repeated again, until we find no maximal k-core.

The authors defined the influence of a community equal to the least influential node present in the community. Due to the use of this definition, the influence value of the community stored at any iteration will be the least among all the communities present in the k-core at that iteration, since it contains the node with the least influence. This ensures that the influence value of communities stored at every iteration keeps on increasing. The final output of the algorithm, i.e. the top-r k-influential communities, are the communities stored in the last r iterations.

Calculating k-core for every iteration was identified by the authors as the costliest part of the algorithm. In a graph with \begin{math}m\end{math} edges, the time taken by the algorithm to perform \begin{math}n_k\end{math} iterations is \begin{math}O(n_km)\end{math} \cite{Li:2015:ICS:2735479.2735484}. It is easy to notice that the only change between the maximal k-core of two consecutive iterations occurs in the community from which the least influential node was removed. Thus, as suggested by Li et al \cite{Li:2015:ICS:2735479.2735484}, we can easily avoid recalculating the k-core at every iteration by using the k-core of the previous iteration, and removing certain nodes from it which are not a part of the k-core anymore. These nodes can be easily identified by performing DFS from the removed node \cite{Li:2015:ICS:2735479.2735484}. The optimization proposed above reduces the time complexity of our algorithm to \begin{math}O(m+n)\end{math}, where \textit{m} and \textit{n} represent the edges and vertices present in the graph respectively.

\textbf{ICPS} : The method of using DFS is sufficiently fast when dealing with only a single query of finding the top-r k-influential communities. However, repeating the same process for different values of k becomes infeasible for large graphs. A naive approach to overcome this could be to store the top-r k-influential communities for every value of k upto some \begin{math}k_{max}\end{math}. The time complexity now will be \begin{math}O(k_{max}(m+n))\end{math} for any number of queries. An obvious limitation to this method is the amount of space required to store the answers for every value of k.

To counter this limitation, Li et al \cite{Li:2017:FIC:3159176.3159197} introduced a novel data structure, called influential community-preserved structure (ICPS). This structure compresses and stores the top-r k-influential communities for all values of k upto \begin{math}k_{max}\end{math} in space complexity \begin{math}O(m)\end{math}. Once this data structure is created, it takes time \begin{math}O(n_a)\end{math}, where \begin{math}n_a\end{math} represents the number of vertices in the answer, to extract the answer for any query. Li et al \cite{Li:2017:FIC:3159176.3159197} also suggested an optimization to this precomputation, reducing the time complexity to \begin{math}O(pm)\end{math}, where p is the arboricity of the graph, m total number of edges. The arboricity of an undirected graph is the minimum number of forests into which its edges can be partitioned.

\textbf{Forward / Backward Algorithm} : Chen et al \cite{chen2016efficient} provided two important extensions to the above algorithm, namely forward and backward algorithms. They identified the computation of Maximal Connected Component (MCC) as the slow step of the algorithm. MCC computation is done at the end of each iteration, to find the parent community of the deleted node. The forward algorithm, maintain a hash table to keep track of the deleted nodes and then computes the MCCs later in the reverse order, thereby ensuring that only the top-r most influential communities are computed.

The backward algorithm, start with the most influential node present in the graph and then adds nodes in decreasing order of influence. After every addition, the algorithm checks whether the k-core property is satisfied, thus finding the top-r k-influential communities from most influential to least. The process is rather slow as compared to forward algorithm, however outperforms it for very small values of r. Thus, for smaller values of r, backward algorithm should be used, and switch to forward algorithm when the value of r increases. Chen et al \cite{chen2016efficient} also used WebGraph \cite{mohapatra2002webgraph}, a graph compression framework to reduce the space complexity of the compressed graph storage.

\textbf{LocalSearch} : Bi et al \cite{Bi:2018:OPA:3213880.3232249} introduced LocalSearch, a novel algorithm that works by reducing the community search problem to a local area in a subgraph containing only the nodes with influence greater than a threshold. The threshold influence is chosen such that it is the minimum influence at which there are atleast r k-influential communities in the subgraph. Any k-influential community not found in this subgraph, but actually present in the original graph, will have influence less than any community found in this subgraph (since it will contain atleast one node with influence less than the threshold). This way, the top-r k-influential communities found in this subgraph will be the top-r k-influential communities of the complete graph.

The algorithm proposed starts with a heuristically chosen threshold and then expands the graph until there are at least r k-influential communities in the sub graph. To speed up the calculation of the number of k-influential communities in a subgraph, the authors define a new term, keynode, which is simply the least influential node of a k-community. The advantage of finding just the keynodes is that we do not need to calculate the MCCs of the graph. While removing least influential nodes from the k-core, as done in the forward algorithm by Chen et al \cite{chen2016efficient}, the algorithm also maintains an array of all the nodes deleted during DFS at each iteration. After all the iterations, instead of calculating MCCs for the top r keynodes, it introduces a different recursive method of identifying the communities. We start with all the nodes deleted at the \begin{math}(T-p)^{th}\end{math} iteration, where T is the total number of iterations. These are extended to include all the nodes in the \begin{math}(p+1)^{th}\end{math} k-influential community which are connected to them, thus resulting in the \begin{math}p^{th}\end{math} k-influential community. The first k-influential community is simply all the nodes deleted in the last iteration. The proof of correctness can be found in \cite{Bi:2018:OPA:3213880.3232249}.

\textbf{kr-clique} : We have been discussing till now the optimizations for the algorithm defined by Li et al \cite{Li:2015:ICS:2735479.2735484}. However, the definition by Li et al \cite{Li:2015:ICS:2735479.2735484} is not universally accepted. Wang et al \cite{unknown} recently introduced a new definition of influence, and focused on the idea of using kr-clique rather than the traditional k-core for finding top-r k-influential communities. We will not be discussing the definition used by them, however will discuss the algorithms they used, since their proposed solutions can be easily translated to our problem statement. A kr-clique is defined such that every node has k edges and every nodes is reachable in atmost r hops from any node. This ensures more cohesiveness in the community than a k-core, since in a k-core two nodes can be very far from each other, which is not true in a kr-clique. 

Wang et al \cite{unknown} first provides a baseline algorithm for their influence model. The algorithm iterates through all the nodes in the graph, computing maximal kr-clique for that node and then calculating its influence. This is clearly a brute force algorithm, not suitable for large graphs. To improve the performance of the algorithm, the authors introduce a novel data structure, namely C-tree, which indexes maximal r-cliques generated by the graph. C-trees are space and time efficient when it comes to generating kr-cliques. The authors list four methods of searching through the C-tree, sequential-order based (SO) search, improved sequential-order based (SO+) search, best-first based (BF) search, and fast best first based (BF+) search. 

\textbf{Progressive Approach} : All the algorithms discussed above outputs the final answer, i.e all the top-r k-influential communities, in one go. This leads to a significant latency between the query and the result. To handle this Bi et al \cite{Bi:2018:OPA:3213880.3232249} suggests a progressive approach in which the communities are generated in decreasing order one after the other. In the first iteration, the problem statement is finding the top-1 k-influential community. The threshold and subgraph is selected, as mentioned in LocalSearch, and the top-1 k-influential community is provided as the output. Now, the threshold is shifted, so as to double the size of the subgraph. The LocalSearch now outputs another set of k-communities. This process is repeated until the require number of k-communities are provided.

\section{Datasets}
Bi et al \cite{Bi:2018:OPA:3213880.3232249} provides an extensive comparison between the algorithms discussed in this survey \cite{Li:2015:ICS:2735479.2735484, Li:2017:FIC:3159176.3159197, Bi:2018:OPA:3213880.3232249, chen2016efficient} by doing experiments across a variety of datasets. All the experimentation results presented here were taken from Bi et al \cite{Bi:2018:OPA:3213880.3232249}. The algorithms compared are,
\begin{enumerate}
\item OnlineAll :- DFS and ICPS based algorithm by Li et al \cite{Li:2015:ICS:2735479.2735484, Li:2017:FIC:3159176.3159197}.
\item Forward Algorithm :- Calculating MCCs for only the relevant communities, by Chen et al \cite{chen2016efficient}.
\item Backward ALgorithm :- Iterating through the most influential nodes in a decreasing order, outperforming forward algorithm for smaller values of r, by Chen et al \cite{chen2016efficient}.
\item LocalSearch :- Local Search Algorithm as presented by Bi et al \cite{Bi:2018:OPA:3213880.3232249}
\item LocalSearch-OA :- Local Search Algorithm presented by Bi et al \cite{Bi:2018:OPA:3213880.3232249}. But using OnlineAll, instead of keynodes. (For comparison purpose)
\item LocalSearch-P :- Local Algorithm algorithm with the progressive approach integrated as shown by Bi et al \cite{Bi:2018:OPA:3213880.3232249}.
\end{enumerate}

\begin{figure}
\includegraphics[width=0.5\textwidth]{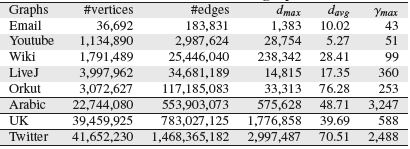}
\caption[]{Statistics of real graph. Image courtesy Bi et al \cite{Bi:2018:OPA:3213880.3232249}}
\label{graph_stats}
\end{figure}

All algorithms were implemented in C++ and compiled in GNU GCC 4.8.2 with the -O3 flag. The experiments were conducted on Intel i5 3.20GHz CPU and 16GB main memory \cite{Bi:2018:OPA:3213880.3232249}.

\textbf{Graphs} : The dataset contains eight real graphs, Email, Youtube, Wiki, LiveJ, Orkut, Arabic, UK and Twitter. The first five are from the Stanford Network Analysis Platform \cite{stanford}, and the last three are from the Laboratory of Web Algorithmics \cite{web}. The variability across different graphs can be found in Fig \ref{graph_stats}. The two parameters r and k were varied as follows, r chosen from \{5, 10, 20, 50, 100\} and k chosen from \{5, 10, 20, 50\}. For every query the algorithms were run three times and the average CPU time in milliseconds was reported. \\

\scriptsize
\textbf{Note} : The running convention used by Bi et al \cite{Bi:2018:OPA:3213880.3232249} is finding the top-k \begin{math}\gamma\end{math}-influential communities, which conflicts with the conventions used in all the papers covered in this survey. We will be sticking to our convention, however the graphs depicted in the later sections, courtesy Bi et al \cite{Bi:2018:OPA:3213880.3232249}, contains the conventions used by them, which should not be correlated with the variables of our convention. Each graph is provided with a descriptive label to avoid any confusion.

\normalsize

\begin{figure*}
\includegraphics[width=1\textwidth]{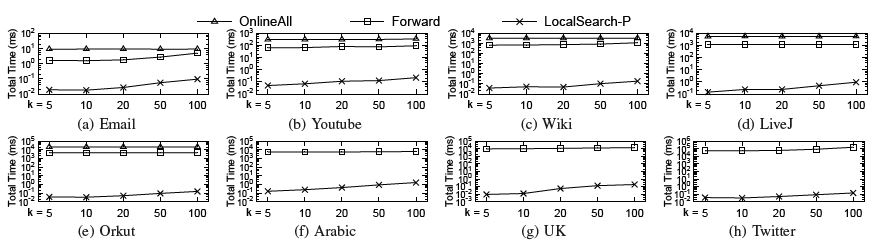}
\caption[]{A brief comparison across all the datasets. Value of r is varied keeping k=10. Courtesy Bi et al \cite{Bi:2018:OPA:3213880.3232249}}
\label{fig2}
\end{figure*}

\begin{figure*}
\includegraphics[width=1\textwidth]{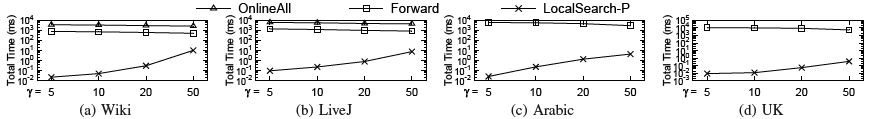}
\caption[]{A brief comparison across all the datasets. Value of k is varied keeping r=10. Courtesy Bi et al \cite{Bi:2018:OPA:3213880.3232249}}
\label{fig3}
\end{figure*}

\begin{figure*}
\includegraphics[width=1\textwidth]{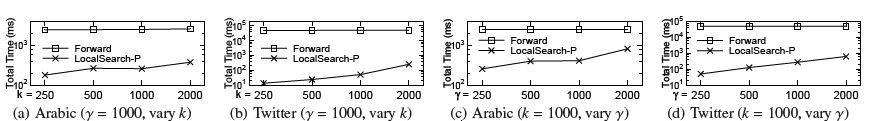}
\caption[]{Comparison of Forward Algorithm and LocalSearch-P for higher values of k and r. Courtesy Bi et al \cite{Bi:2018:OPA:3213880.3232249}}
\label{fig4}
\end{figure*}

\begin{figure*}
\includegraphics[width=1\textwidth]{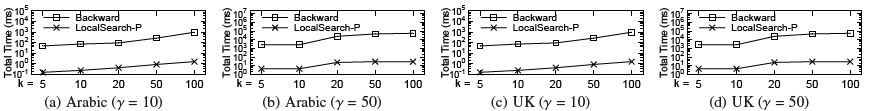}
\caption[]{Comparison of Backward Algorithm and LocalSearch-P for varying values of k and r. Courtesy Bi et al \cite{Bi:2018:OPA:3213880.3232249}}
\label{fig5}
\end{figure*}

\begin{figure*}
\includegraphics[width=1\textwidth]{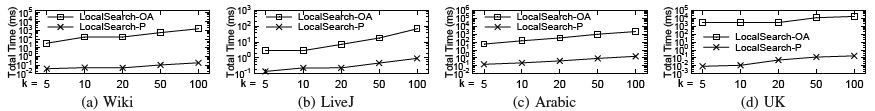}
\caption[]{Comparison between LocalSearch-OA and LocalSearch-P for varying values of r, keeping k = 10. Courtesy Bi et al \cite{Bi:2018:OPA:3213880.3232249}}
\label{fig6}
\end{figure*}

\begin{figure*}
\includegraphics[width=1\textwidth]{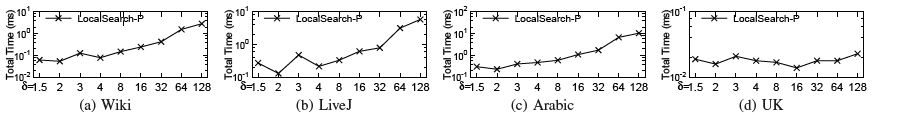}
\caption[]{Comparison of different of values of the ratio at which the graph is grown in LocalSearch-P, for k=10 and r=10. Courtesy Bi et al \cite{Bi:2018:OPA:3213880.3232249}}
\label{fig7}
\end{figure*}

\begin{figure*}
\includegraphics[width=1\textwidth]{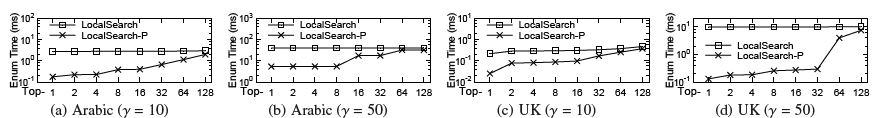}
\caption[]{Comparison between LocalSearch and LocalSearch-P for r=128, k=10 and 50. Courtesy Bi et al \cite{Bi:2018:OPA:3213880.3232249}}
\label{fig8}
\end{figure*}

\section{Comparative Analysis}
The OnlineAll and Forward algorithms work on the whole graph for any value of k and r and hence there is very little change in the total time taken for processing queries when the values of k or r are varied. The forward algorithm performs slightly better than the OnlineAll algorithm because of the reduced number of Maximal Connected Component(MCC) computations. (refer to Fig \ref{fig2},\ref{fig3},\ref{fig4})

The Backward algorithm starts from the most influential community and stops when the top-r most influential communities have been found. Hence the time taken by Backward algorithm increases with the increase in the value of r (refer to Fig \ref{fig5}).

The total time for LocalSearch algorithm increases as we increase r keeping k constant. This can be easily explained, since at a higher value of r, the subgraph formed will be larger in size and hence the time taken will be more. Similar reasoning can be given for varying k and keeping r constant. The increase in the value of k means a larger sub graph is required to satisfy the stronger cohesive property and hence an increase in total time can be seen (refer to Fig \ref{fig8}).

LocalSearch-P outperforms every algorithm that we have discussed in this paper, including the variations LocalSearch and LocalSearch-OA (refer to Fig \ref{fig2},\ref{fig3},\ref{fig4},\ref{fig5},\ref{fig6},\ref{fig8}).

A comparison was done for the time taken to output  top-p k-influential communities, p varying for 0 to r. Since LocalSearch gives the complete output in one go, the time for every value of p is the same. However, the progressive approach delivers the initial results almost immediately and then the time taken gradually increases (refer to Fig \ref{fig8}).

\section{Extensions}
In this section we will discuss a few extensions to the problem statement as discussed in this survey.

\subsection{Non-containment Community Search}
Following the definition of k-communities, it is possible that an influential k-community is a subgraph of another influential k-community. However, for most of the practical applications, one would prefer to not have such relations present in the output. Thus the problem statement of finding top-r non-containment influential k-communities was introduced. 

OnlineAll \cite{Li:2015:ICS:2735479.2735484} checks for every community generated whether it contains a k-core or not and only those are considered for the final output which have no k-cores present in it. Forward Algorithm \cite{chen2016efficient} uses an additional array to store the degree of a node and then used its value to check for non-containment. A slight modification in LocalSearch \cite{Bi:2018:OPA:3213880.3232249} was also suggested in the algorithm to tackle the above task. To summarize we can say that the modified problem statement can be easily solved by providing slight modifications to the existing algorithms.

\subsection{Other Cohesiveness Measures}
In this survey we came across two structures which were used to define cohesiveness in a graph, the k-core definition \cite{chen2016efficient,Li:2015:ICS:2735479.2735484,Bi:2018:OPA:3213880.3232249} to ensure that each node is connected to at least k different nodes and kr-clique \cite{unknown} to concentrate more on the compactness in a graph. The kr-clique structure puts a limit to the maximum distance between two nodes in a community thereby ensuring more compactness. This also results in lesser number of k-communities. The definition of cohesiveness can depend on our need or on the problem statement we are trying to solve and different algorithms may be experimented with to get the best results.

\subsection{Mutli-dimensional Influence value}
Till now in our discussion, the influence of a node is a single numerical attribute associated with it. For many practical purposes however, one single numerical value might not be enough to represent the influence of a node. For example, in a collaboration network every author can have multiple attributes like number of published papers, sociability, diversity, activity etc. The simplest approach might be to take a linear function of these attributes as the one numerical attribute and use the algorithms discussed above but this might not capture the details as desired. A novel community model, named the skyline community model, was introduced recently Li et al \cite{Li:2018:SCS:3183713.3183736} to solve this specific problem.

\section{Conclusion and Future Work}
We have produced a survey that can be used to get familiar with the domain of influential community search and get the reader acquainted with the research already done in this field. We provide an analysis of the basic k-core algorithm used and several optimization proposed over the years. In this survey we cover several methods for searching the top-r k-influential communities, comparing various optimizations, and touching on the varying definition of influence. 

One possible direction of future work can be exploring even more definitions of cohesiveness in order to obtain tightly knit communities. The other possible direction of future work is to explore the possibility of updating of influence of nodes online and to integrate the existing algorithms into compressed graphs that fit in the memory. These are all interesting areas of future work.

\bibliographystyle{unsrt}

\end{document}